\documentclass[aps,prl,amsmath,superscriptaddress,twocolumn,longbibliography]{revtex4-2}

\usepackage{graphicx}
\usepackage{amsmath}
\usepackage{amsfonts}
\usepackage{amssymb}
\usepackage{braket}
\usepackage{bm}
\usepackage{multirow}
\usepackage{color}
\usepackage[normalem]{ulem}
\usepackage{mathrsfs}
\usepackage{mathtools}
\usepackage{float}
\usepackage{booktabs}
\usepackage{amsmath}

\makeatletter

\newcommand{\Rmnum}[1]{\expandafter\@slowromancap\romannumeral #1@}
\makeatother

\usepackage[breaklinks]{hyperref}
\hypersetup{colorlinks=true, linkcolor=blue, citecolor=blue, filecolor=blue, urlcolor=blue}

\AtBeginDocument{%
    \newwrite\bibnotes
    \def\bibnotesext{Notes.bib}
    \immediate\openout\bibnotes=\jobname\bibnotesext
    \immediate\write\bibnotes{@CONTROL{REVTEX41Control}}
    \immediate\write\bibnotes{@CONTROL{%
    apsrev41Control,author="08",editor="1",pages="1",title="1",year="1"}}
     \if@filesw
     \immediate\write\@auxout{\string\citation{apsrev41Control}}%
    \fi
}%

\begin{document}

\title{Topological Electronic and phononic chiral edge states in  SiTc Crystal}

\author{Shivendra Kumar Gupta}
\email{shivendrakumarg900@gmail.com}
\affiliation{Department of Physics, Visvesvaraya National Institute of Technology, Nagpur, 440010, India \\}

\author{Saurabh Kumar Sen}%
 \author{Nagarjuna Patra}%
 \author{Ajit Singh Jhala}%
\author{Poorva Singh}
\email{poorvasingh@phy.vnit.ac.in}
\affiliation{Department of Physics, Visvesvaraya National Institute of Technology, Nagpur, 440010, India \\}

\date{\today}

\begin{abstract}
   Topological materials hosting multifold fermions and bosons have emerged as a rich platform for exploring unconventional quasiparticles and transport phenomena. In this work, we investigate the chiral crystal SiTc using first-principles density functional theory and symmetry-based analysis to explore its topological electronic and phononic properties. Our study identifies multiple high-fold degeneracies and topological nodes in both the electronic band structure and phonon dispersion. We have analyzed the impact of spin–orbit coupling on the evolution of band crossing and identified Weyl points and their associated chiralities. Surface electronic states, Fermi arcs, Berry curvature distributions, and intrinsic spin Hall conductivity are computed to probe the topological response. On the phononic side, we uncover topologically nontrivial bosonic modes and corresponding longest possible Fermi arc features. These results establish SiTc as a promising candidate that simultaneously hosts topological fermionic and bosonic excitations, offering new opportunities for investigating the interplay between electronic and phononic topology.

\end{abstract}

\maketitle

\section{ Introduction}

The discovery of novel fermionic quasiparticles in condensed matter systems have unveiled deep connections between high-energy physics and solid-state quantum materials, paving the way for groundbreaking advancements in both fields. From massless Dirac fermions in graphene\cite{castro2009electronic,young2015dirac} to Majorana quasiparticles in superconducting heterostructures\cite{fu2008superconducting, oreg2010helical, zhang2013majorana, science.1222360} and Weyl fermions in noncentrosymmetric crystals\cite{yang2015weyl, weng2015weyl, li2023emergence, chang2018magnetic}, condensed matter physics has emerged as a powerful platform for realizing relativistic free fermions within a crystalline framework\cite{bradlyn2016beyond}. This progress is driven by the integration of symmetry analysis, group theory, and crystallographic considerations, providing a comprehensive foundation for understanding the underlying physics.

Currently, research is increasingly focused on chiral crystals, which, by lacking of inversion, mirror, and rotoinversion symmetries, offer a unique platform for hosting multifold fermions in electronic and multifold bosons in phononic systems\cite{sadhukhan2021electronic,hagiwara2024orbital}. These exotic quasiparticles, with no direct analogy in high-energy physics, exhibit unconventional electronic dispersion and robust surface phenomena, such as longest possible topologically protected open Fermi arcs connecting high order opposite chiral node\cite{chang2017unconventional}.

Multifold fermionic and bosonic systems exhibit remarkable applications across various domains of condensed matter physics. In electronic systems, topological semimetals (TSMs) with high Chern numbers act as a source of large spin Berry curvature (SBC), resulting in an enhanced spin Hall effect (SHE). Chiral topological semimetals, hosting Kramers-Weyl points (KWPs) with the highest possible Chern number, give rise to exotic phenomena such as extended Fermi arcs, quantized photogalvanic currents, and magnetoelectric effects\cite{ascencio2023enhanced,chang2018topological,shekhar2018chirality,chang2017unconventional}. These properties make them promising candidates for spintronics, high-efficiency memory devices, next-generation computing technologies and best catalyst for hydrogen evolution reaction(HER)\cite{yang2020topological,zhan2024design,gupta2025topological}. On the other hand, topological phononics (TPs) utilize lattice vibrations to manifest bulk-surface correspondence and topologically protected surface states, which remain robust against backscattering. Additionally, TPs support critical physical phenomena such as phononic quantum anomalous Hall-like effect, the phononic valley Hall effect,  and phononic quantum spin Hall-like effect controlled by multiple-valued degrees of freedom. These effects enhance the potential of TPs in phonon-based waveguides, abnormal heat transport, and acoustic metamaterials\cite{ozawa2019topological,li2021computation}. The coexistence of fermionic and bosonic excitations within topological materials may provides a crucial advantage in thermoelectric\cite{singh2018topological} and superconducting\cite{dong2022superconductivity,mardanya2024unconventional,reddy2024coexistent}  applications as well as speciﬁc heat. The interplay between electronic and phononic topological states suppresses phonon-mediated thermal conductivity while maintaining high electrical conductivity, significantly improving the thermoelectric figure of merit (zT)\cite{singh2018topological}. Furthermore, this interaction enhances the pairing mechanism in unconventional superconductors, leading to superior quantum transport properties\cite{dong2022superconductivity}. Hence, the synergy between electronic and phononic topological excitations required a establish  robust platform for advanced functional materials with exceptional transport characteristics.


Multifold fermions extend beyond the traditional Dirac and Weyl fermions. They arise at high-symmetry points in the Brillouin zone, where band crossings create multifold degeneracy. These crossings are protected by the crystal symmetries of the material. Weyl fermion, a smallest well known fermion also called spin 1/2 fermions or two-fold fermions can be present by broken time-reversal symmetry or inversion symmetry or  both\cite{burkov2016topological}. These Weyl fermions have pair of Weyl points of opposite chirality and are connected by open Fermi arc.  Several space groups, including 198 host such multifold fermions, due to their non-symmorphic and chiral lattice properties. Spin-1, charge-2 Dirac fermions, and spin-3/2 chiral fermions are classifications of multifold fermions, represented by Hamiltonian $H \propto \hbar \delta k .L$. Where L is spin matrices of a particular spin (L = $\sigma_{2\times2}$  Pauli spin matrices for spin-1/2, $L_{3\times3}$ for spin-1, and  $L_{4\times4}$ for spin-3/2 systems). The spin-1/2 doublet with chern number $\pm$2 can have symmetry-protected four-fold degeneracy and can be defined as $H \propto \hbar \delta k .\sigma \oplus \mathbf \mathbb{I}_{2\times2} $ also known as charge-2 Dirac node\cite{zhang2018double,barman2020symmetry}.

In this work, we have studied electronic and phononic properties of SiTc material, that have already been synthesized and possesses cubic crystal symmetry (space group 198). SiTc is also been observed as a topological catalyst by using it large non-trivial energy window and show excellent enhancement in hydrogen evolution reaction process\cite{zhan2024design}. We provide a detailed symmetry analysis with first-principles density functional theory calculations along with symmetry protection arguments for the multi-fold system applicable for both electronic and phononic properties. The discussion is based on the Kramers theorem and the commutation and anti-commutation of the generators that relates it to density functional theory results confirming SiTc material possesses multi-fold fermionic systems at the $\Gamma$ and R high symmetry point, neutralized by some Weyl points and has chirality to be zero for a complete Brillouin zone. When spin–orbit coupling (SOC) is not considered, the $\Gamma$ point is three-fold degenerate, while the R point is four-fold degenerate, near the Fermi level in both electronic and phononic band structure. To see the effect of SOC in the electronic properties of SiTc, we perform SOC induced calculation that results in modifying the multifold nature,  the $\Gamma$-point becomes a four-fold degenerate, while the R-point becomes a six-fold degenerate. These results have been supported by the surface state, Fermi-arc, Weyl nodes, and Berry curvature calculations.

\section{Computational Methods}
Electronic and phononic properties of SiTc compound has been  investigated via first-principles calculation based on standard density functional theory\cite{kohn1965self} with the potential linearized augmented plane-wave method provided by the VASP package\cite{hafner2008ab, blochl1994projector, kresse1996efficient}. Generalized gradient approximation (GGA) with projector augmented wave (PAW)\cite{torrent2010electronic} potentials have been utilized to incorporate the exchange-correlation function. The plane-wave basis energy cutoff of 500 eV and $\Gamma$-centered Monkhorst-pack\cite{monkhorst1976special} k-grid of $11 \times 11 \times 11$ has been used to perform self-consistent calculations (SCF). The band structures, excluding and including of SOC, were calculated using optimized lattice parameter. Maximally localized Wannier functions (MLWF) are used to develop the tight-binding model, which is used to calculate the surface states of SiTc material using the Wannier90 code \cite{mostofi2008wannier90}. Wannier Tools is used to obtain topological characteristics such as topological chiral surface states and Fermi arcs\cite{wu2018wanniertools}. To study the force constant and phonon dispersion of SiTc, we preformed density functional perturbation theory calculation with $3\times  3\times 3$ supercell and energy convergence of $10^{-8}$ eV\cite{togo2015first}. A tight binding hemiltonian has been created with the PHONOPYTB tool \cite{phonopyTB} and wannier tool package is used to perform topological phonons properties\cite{wu2018wanniertools}.

According to the Kubo formalism, the integration of spin Berry curvature across the entire Brillouin zone, summed over all occupied electronic states, leads to the intrinsic spin Hall conductivity\cite{guo2008intrinsic,guo2005ab}.
The intrinsic spin Hall conductivity (SHC) is given by:
\begin{equation}
\sigma^{k}_{ij} = e \hbar \sum_n \int_{\mathrm{BZ}} \frac{d\mathbf{k}}{(2\pi)^3} f_{n\mathbf{k}} \, \Omega^{s,k}_{n,ij}(\mathbf{k}),
\end{equation}
where $\Omega^{s,k}_{n,ij}(\mathbf{k})$ is spin Berry curvature (SBC) for $n$-th band at wavevector $\mathbf{k}$.

The spin Berry curvature is defined as:
\begin{equation}
\Omega^{s,k}_{n,ij}(\mathbf{k}) = - \sum_{n' \ne n} \frac{2 \, \mathrm{Im} \left[ \bra{n\mathbf{k}} \hat{J}^k_i \ket{n'\mathbf{k}} \bra{n'\mathbf{k}} \hat{v}_j \ket{n\mathbf{k}} \right]}{(\epsilon_{n\mathbf{k}} - \epsilon_{n'\mathbf{k}})^2},
\end{equation}

Here, the spin current operator is defined as $\hat{J}^k_i = \frac{1}{2} { \hat{v}_i, \hat{s}_k }$, where $\hat{v}_i$ denotes the velocity operator and $\hat{s}_k = \frac{\hbar}{2} \hat{\sigma}_k$ is spin operator expressed in terms of the Pauli matrices $\hat{\sigma}_k$. The indices $i$, $j$, and $k$ correspond to the cartesian coordinates ($x$, $y$, $z$). To achieve numerical accuracy in evaluating the spin Berry curvature over the whole Brillouin zone (BZ), a dense $\mathbf{k}$-point mesh of $125 \times 125 \times 125$ was employed.

\section{Symmetry Analysis}

The presence of multifold fermions is restricted to primitive cubic, primitive tetragonal, or body-centered cubic lattices with non-symmorphic space groups, where inversion symmetry is broken. In space group 198, the crystal structure exhibits tetrahedral (\(T_d\)) point-group symmetry, which plays a crucial role in the symmetry analysis at high-symmetry points in the Brillouin zone\cite{bradlyn2016beyond}. 


At the \(\Gamma\) point, the little group coincides with the point group and is generated by two screw symmetries, 
\[
S_{2z} = \{C_{2z} \mid \frac{1}{2}, 0, \frac{1}{2}\}, \quad S_{2y} = \{C_{2y} \mid 0, \frac{1}{2}, \frac{1}{2}\}, 
\]
along with a three-fold rotation, 
\[
S_3 = \{C_3^+, 111 \mid 0, 0, 0\}, 
\]
about the (111) diagonal of the unit cell. These generators ensure the existence of a three-dimensional irreducible representation (irrep), and satisfy these following two relations that are only applicable for cubic system\cite{barman2020symmetry}.
\[
S_{2z} S_{3} = S_{3} S_{2y}, \quad S_{3} S_{2z} S_{2y} = S_{2y} S_{3}.
\]

For non-magnetic materials, time-reversal symmetry (\(\mathcal{T}\)) is preserved, leading to the constraint \( \mathcal{T}^2 = I \). The screw symmetries \( S_{2y} \) and \( S_{2z} \) commute, satisfying
\[
[S_{2y}, S_{2z}] = 0, \quad S_{2y}^2 = I, \quad S_{2z}^2 = I.
\]
The three-fold rotation \( S_3 \) plays a crucial role in determining band degeneracies, as it acts nontrivially on eigenstates, leading to the formation of three-fold degeneracies when the representation is irreducible. If \( S_3 \) acts trivially, the system instead hosts two-fold or non-degenerate states. Consequently, the interplay between these symmetry elements dictates the spin-1 excitation of electronic bands at \(\Gamma\), paving the way for the emergence of multifold fermionic excitations in these systems\cite{barman2020symmetry, bradlyn2016beyond}.


At the \( R \)-point, the little group consists of three symmetry generators, including two screw symmetries,  
\[
S_{2x} = \{C_{2x} \mid 0, \frac{3}{2}, \frac{1}{2}\}, \quad S_{2y} = \{C_{2y} \mid 0, \frac{3}{2}, \frac{1}{2}\},
\]
and a three-fold anticlockwise rotation about the (111) diagonal of the unit cell,  
\[
S_{3} = \{C_3^-, 111 \mid 0, 1, 0\}.
\]  
These generators also satisfy the relations \( S_{2x} S_{3} = S_{3} S_{2y} \) and \( S_{3} S_{2x} S_{2y} = S_{2y} S_{3} \), which determine the structure of degeneracies at the \( R \)-point.  A key distinction at \( R \) arises due to the screw symmetries \( S_{2x} \) and \( S_{2y} \), which anticommute, satisfying \( \{S_{2x}, S_{2y}\} = 0 \). Additionally, their squares satisfy \( S_{2x}^2 = -I \) and \( S_{2y}^2 = -I \), leading to a fundamental constraint on the degeneracies of electronic states. In the $P2_13$ (No. 198) space group of chiral crystals, the $R$-point is a time-reversal invariant momentum (TRIM). The associated screw rotation symmetry must commute with the time-reversal symmetry (TRS) operator, ensuring that all symmetry representation eigenvalues are real. However, the condition is not satisfied for a two-dimensional representation of $SU(2)$ but can be fulfilled by a four-dimensional representation with charge $\pm2$.

The two-fold degeneracy originates from the anticommutation of screw symmetries, where an eigenstate \( |\psi\rangle \) and its transformed counterpart \( S_{2x}|\psi\rangle \) form distinct eigenstates of \( S_{2y} \) with eigenvalues \( i \) and \( -i \), respectively. An additional two-fold degeneracy is introduced through the three-fold rotation \( S_3 \), where \( S_3|\psi\rangle \) and \( S_{2y} S_3|\psi\rangle \) serve as distinct eigenstates of \( S_{2x} \) with eigenvalues \( i \) and \( -i \), respectively. To ensure the mutual orthogonality of these states, \( S_3 \) must act nontrivially, meaning \( \langle\psi|S_3|\psi\rangle \neq 0 \). Since no additional constraints generate any new states for spinless system no higher than four-fold is possible at R point. So ultimately, the combination of screw symmetry anticommutation, three-fold rotation, and time-reversal constraints results in a symmetry-protected four-fold degeneracy at \( R \), confirming the presence of an exactly four-fold node dictated by the crystalline symmetries throughout the energy range\cite{barman2020symmetry, bradlyn2016beyond}.  

\begin{figure}[!t]
	 \includegraphics[width=.8 \linewidth]{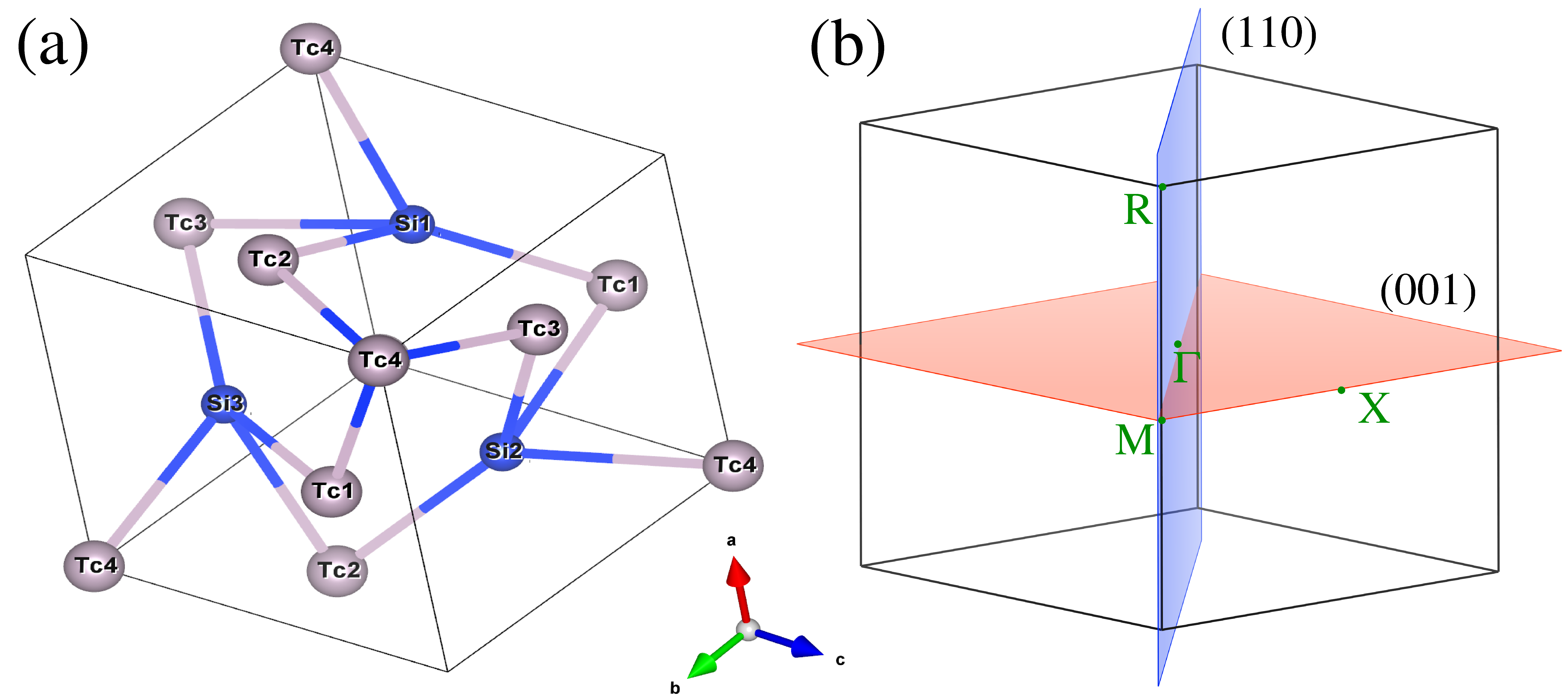}
	 \caption{ (a) Crystal structure of SiTc displayed in its conventional cubic unit cell configuration, highlighting the atomic arrangement and symmetry. (b) Corresponding bulk and projected surface Brillouin zones of SiTc, illustrating high-symmetry points critical for electronic band structure calculations and surface state analysis.}
	  \label{fig:figure1}
\end{figure}



In spinful systems, the presence of time-reversal symmetry imposes the constraint $T^2 = -I$, which leads to Kramers degeneracy. However, this degeneracy is only guaranteed at time-reversal invariant momenta, limiting its influence at general points in the Brillouin zone. The inclusion of spin-orbit interaction in systems lacking inversion symmetry significantly impacts the degeneracy structure of electronic states. Specifically, at the $\Gamma$-point, a potential three-fold spinless degeneracy cannot evolve into a six-fold spinful degeneracy upon the inclusion of spin-orbit coupling. Instead, the states split into two distinct groups: a four-fold degenerate state and a two-fold degenerate state, dictated by symmetry constraints.

The reason behind this restricted degeneracy structure lies in the symmetry constraints imposed at $\Gamma$. The screw symmetries $S_{2y}$ and $S_{2z}$ anticommute and square to $-I$, ensuring that only a four-fold degeneracy is protected at this point. This degeneracy arises due to the interplay between point-group symmetries and the eigenstates associated with $S_{2y}$ and $S_{2z}$. Additionally, time-reversal symmetry does not necessarily enforce mutual orthogonality between screw eigenstates and their time-reversed counterparts due to the imaginary nature of their eigenvalues.

More generally, different initial degeneracy conditions lead to distinct outcomes upon incorporating spin-orbit coupling. A singlefold spinless band transforms into a Kramers doublet when SOC is included. Similarly, a spinless two-fold degeneracy evolves into a four-fold spinful degeneracy, consistent with the preserved symmetries. Notably, the behavior of degeneracies at the $R$-point differs from that at $\Gamma$, as six-fold degeneracies are permitted at $R$, whereas such configurations are not allowed at $\Gamma$. This distinction arises from the unique symmetry relations governing each individual high-symmetry point, further emphasizing the intricate role of spin-orbit coupling in shaping the electronic band structure of these systems.

\begin{figure}[!t]
\centering
	 \includegraphics[width=1 \linewidth]{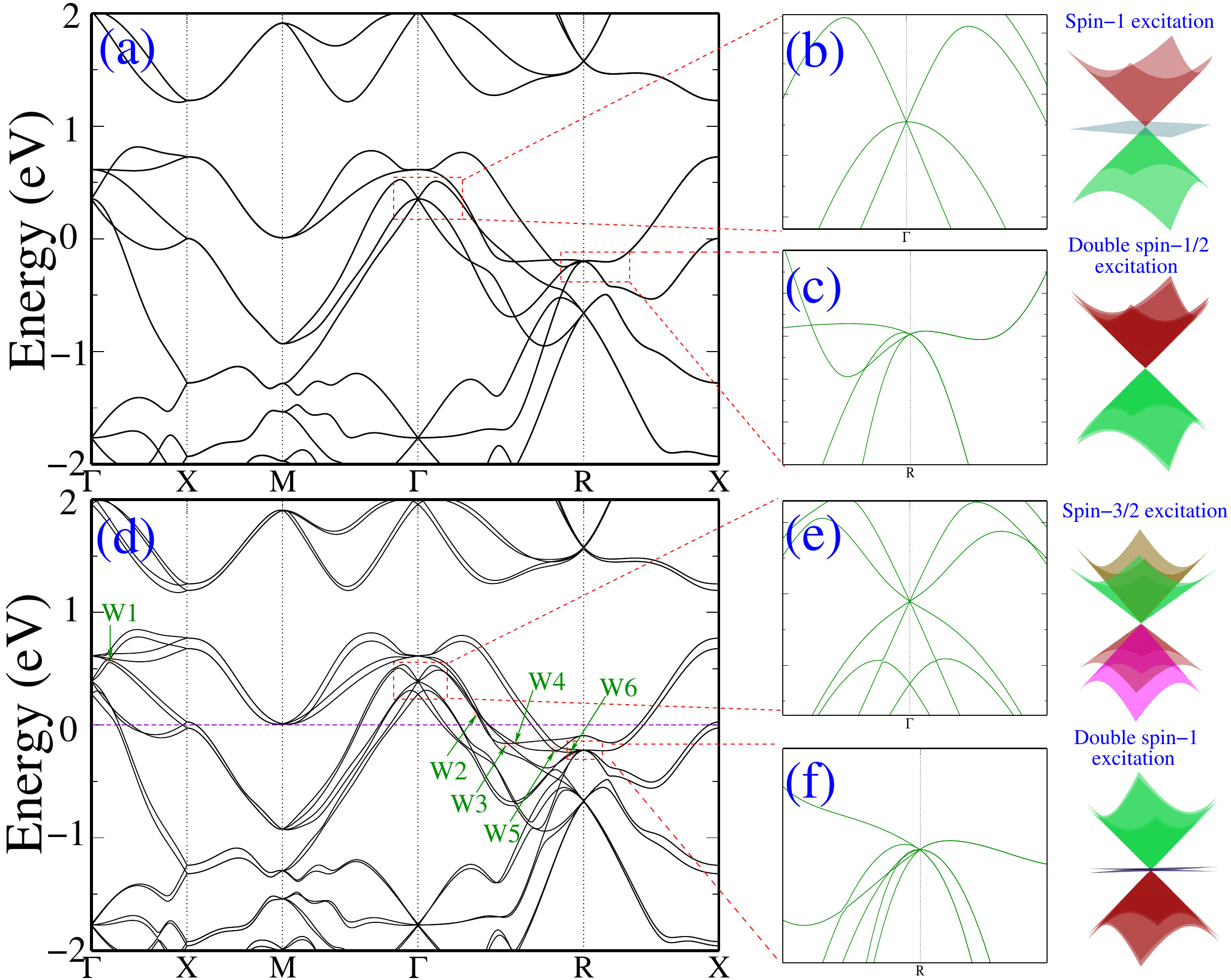}
   \caption{\label{fig:figure2} Electronic band structure of SiTc: (a) without spin-orbit coupling (SOC), (b) enlarged view at the $\Gamma$ point showing a three-fold band crossing, (c) enlarged view at the R point illustrating a four-fold crossing, (d) with SOC, (e) enlarged view at the $\Gamma$ point revealing a four-fold degenerate crossing, and (f) enlarged view at the R point showing a six-fold band crossing.}

\end{figure}

\section{Result and Discussion} 

The chiral crystal material SiTc, which has been experimentally synthesized and belongs to the $P2_13$ (No. 198) space group with a simple cubic structure, has been investigated. The optimized lattice parameter a=4.78$\AA$ is well matched with the experimental results (4.755$\AA$)\cite{schob1964structure} and has greater effecient catalyst than pericious metal Pt, for hydrogen evolution reaction\cite{zhan2024design,gupta2025topological}. The crystal structure and its corresponding bulk and surface Brillouin zone of SiTc has been shown in Figure \ref{fig:figure1}. The electronic band structure of SiTc without and with spin-orbit coupling have been shown in Figure \ref{fig:figure2}(a) and \ref{fig:figure2}(d) respectively. The presence of 3-fold spin-1 excitation fermion crossing at $\Gamma$ and a four-fold double spin-1/2 fermion crossing at R, have been zoomed in Figure \ref{fig:figure1}(b) and \ref{fig:figure2}(c) respectively
 for without SOC case. The symmetrically allowed degeneracy is upto three-fold for  $\Gamma$  however only four-fold degeneracy is  allowed for R high symmetry point.

 \begin{figure}[t]
\centering
	 \includegraphics[width=1 \linewidth]{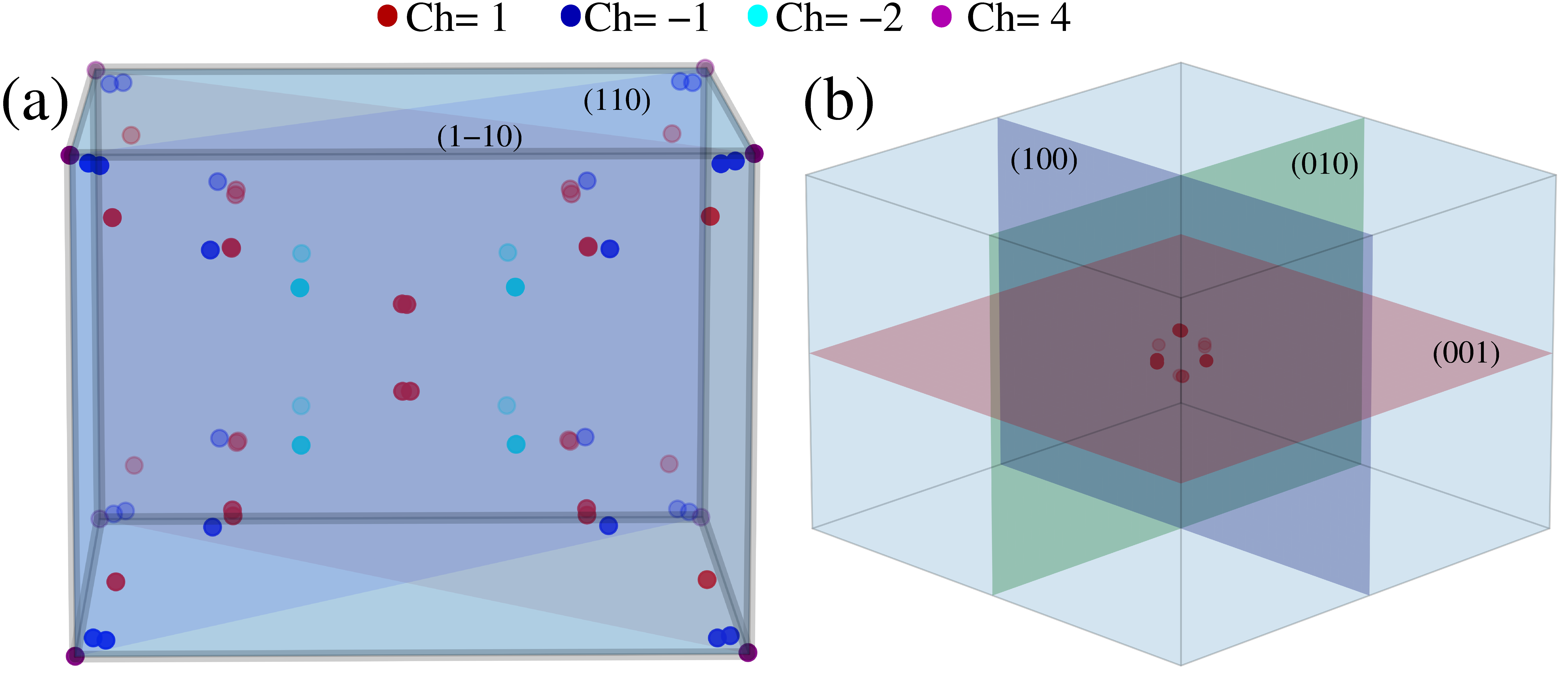}    \caption{\label{fig:figure3} Distribution of Weyl nodes in the first Brillouin zone: (a) Projection on the (110) and (1$\bar{1}$0) planes, and (b) projection on the (100), (010), and (001) planes. Four Weyl points, common to both (a) and (b), are located at the intersection of the (110), (1$\bar{1}$0), and (100) planes, all four possessing a chirality of +1.}
\end{figure}

The inculsion of SOC in the band structure, As the inversion symmetry is absence and Time reveral symmetry is preserved, resulting the splitting of bands between high symmery points however at high symmerty points bands may be degenarte as in Figure \ref{fig:figure2}(d). However we are intrested at multifold fermions present at high symmetry points. The 3-fold of WSOC crossing present at $\Gamma$ split and converted into 4-fold  and a 2-fold crossing due to symmetry allowed upmost 4-fold. Whereas 4-fold WSOC crossing at R split and converted into 6-fold  and a 2-fold crossing due to symmetry allowed upmost 6-fold as discussed in the section II.

From a topological perspective, a four-fold degenerate point is observed at the $\Gamma$ point, located slightly above the Fermi level. However, since all four intersecting bands at this point originate from the valence band manifold, the degeneracy at $\Gamma$ does not carry any net topological chirality. In contrast, the topologically nontrivial band crossings occur between the conduction and valence bands along the high-symmetry paths $\Gamma$–X and $\Gamma$–R within the Brillouin zone. These crossing points, labeled W1 to W6 in Figure \ref{fig:figure2}(d), correspond to Weyl nodes with chiralities of $\pm1$ and $\pm2$. Additionally, a six-fold degenerate point is present at the R point on the Brillouin zone boundary, carrying a net chirality of +4.

To validate the presence and nature of these Weyl nodes, detailed Weyl node calculations were performed. The results reveal that W1 consists of 12 nodes with chirality +1, W2 has 8 nodes with chirality –2, W3 contains 16 nodes with chirality +1, W4 consists of 8 nodes with chirality –1, W5 has 8 nodes with chirality +1, W6 comprises 16 nodes with chirality –1, and the R point includes a single six-fold node with chirality +4. The total chiral charge within the first Brillouin zone sums to zero, in agreement with the Nielsen–Ninomiya no-go theorem, which requires that Weyl nodes appear in pairs of opposite chirality, ensuring topological charge neutrality across the Brillouin zone.

\begin{figure}[t]
\centering
	 \includegraphics[width=1 \linewidth]{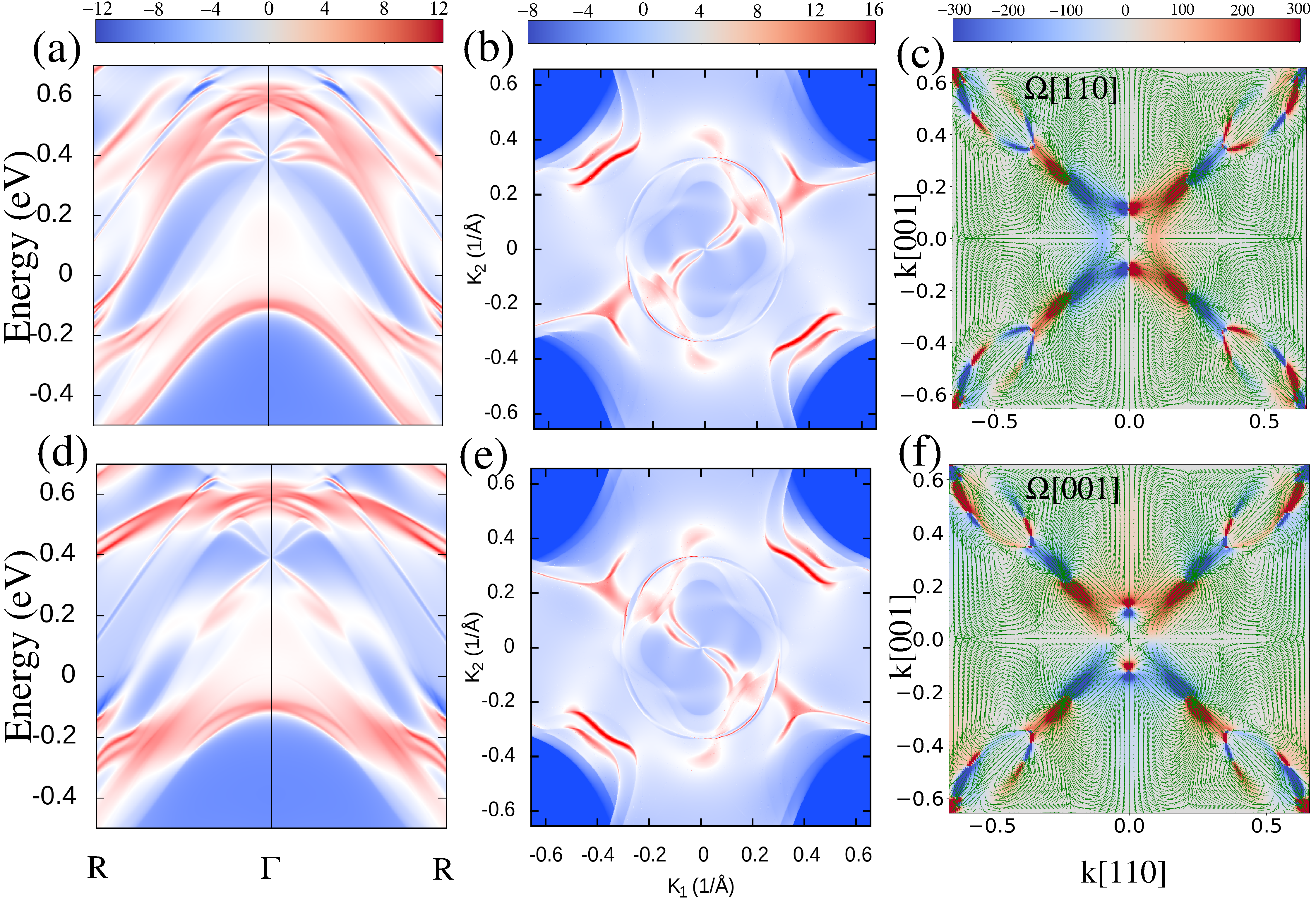}
  \caption{\label{fig:figure4} (a) Surface electronic band structure of the upper (001) surface, and (b) the corresponding Fermi surface. (c) Distribution of Berry curvature $\Omega$ projected along the [110] direction, overlaid with the intensity of its in-plane (xy) and out-of-plane (z) components. Blue and red junctions indicate the presence of Weyl points. (d) Surface electronic band structure of the lower (001) surface, and (e) the corresponding Fermi surface. (f) Distribution of Berry curvature $\Omega$ along the [001] direction, overlaid with the intensity of the xy and z components.}
\end{figure}

To confirm the topological electronic properties of SiTc, we have calculated the surface band structure of 001 upper and lower surfaces (Figure \ref{fig:figure4}(a) and \ref{fig:figure4}(d)) with surface state present and corrosponding Fermi arc (Figure \ref{fig:figure4}(b) and \ref{fig:figure4}(e)) confirming complete chiral open Fermi arc. It can be seen that, the Fermi arc are originating from above and below Weyl points of the $\Gamma$ point and not exectly from $\Gamma$ point. Also we have calculated the distribution of Berry curvature $\Omega$ overlaped with  intensity of
the xy and z components of $\Omega$, shows the local vector fields and the junction of red (positive) and blue (negative) indicating the presence of Weyl point in Figure \ref{fig:figure4}(c) for $\Omega$[110] and \ref{fig:figure4}(e) for $\Omega$[001], that are completely alined with the Weyl nodes calculation in Figure \ref{fig:figure3}.
 
\begin{figure}[!t]
\includegraphics[width=9cm,height=6.2cm]{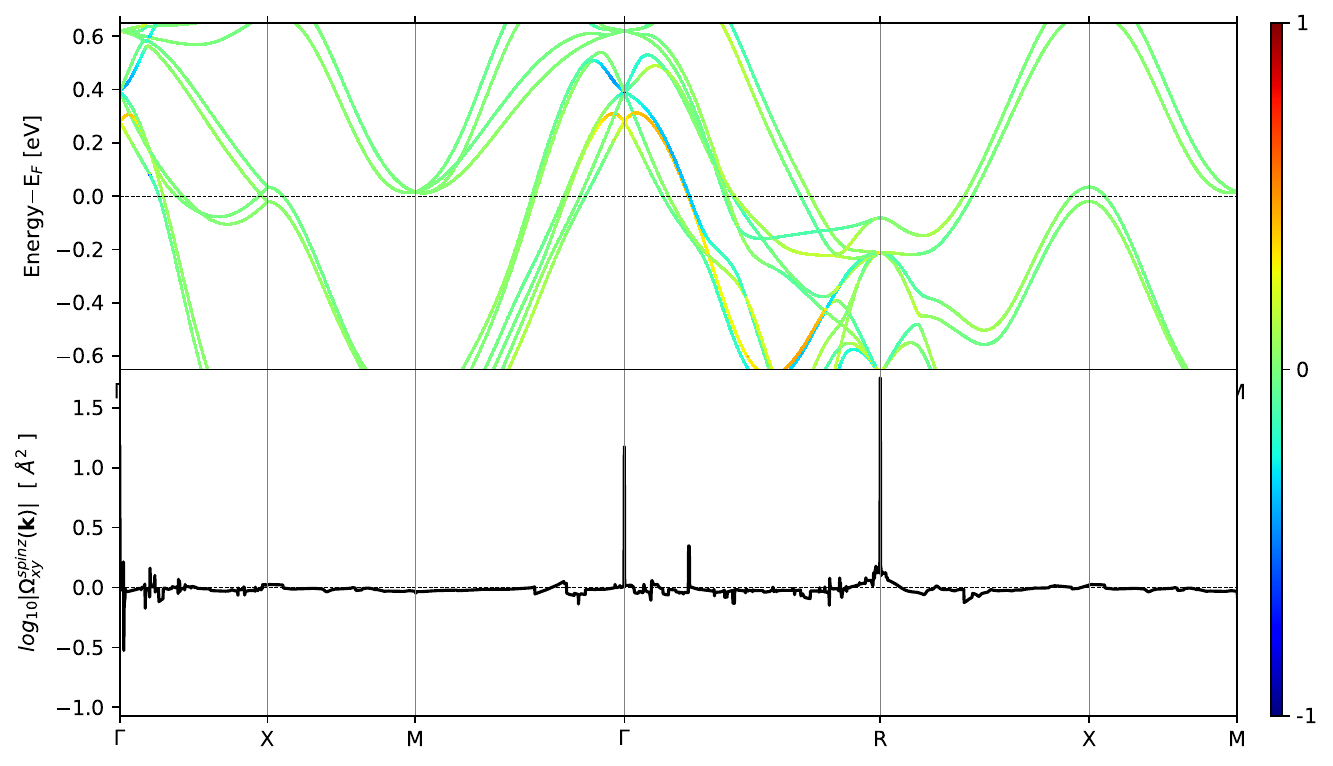}
 \caption{ \label{fig:figure7}  First-principles calculations of the electronic band structure (top panel) and the Berry curvature distribution (bottom panel) of SiTc.}
 \end{figure}

The intrinsic spin Hall conductivity (SHC) serves as a key physical parameter for quantifying the strength of the intrinsic spin Hall effect (SHE) in a material. As a third-rank tensor, the SHC formally comprises 27 components. However, in the case of \textit{SiTc}, the crystal symmetry imposes stringent constraints, significantly reducing the number of independent non-zero elements to only two: $\sigma_{xy}^{z}$ and $\sigma_{xz}^{y}$. To uncover the origin of the pronounced spin Hall conductivity (SHC) observed in \textit{SiTc}, we conducted a detailed analysis of the logarithm of the momentum-resolved spin Hall conductivity component, $\log[\sigma_{xy}^z (\mathbf{k})]$, along the high-symmetry paths at the Fermi energy ($E_F$), as shown in Figure \ref{fig:figure7}. The distribution exhibits a strong dependence on the crystal momentum $\mathbf{k}$, with pronounced peaks near the multifold degenerate points and Weyl nodes along the $\Gamma$–R direction. These features highlight the dominant contribution of these regions to the intrinsic SHC. Furthermore, we have computed the numerical values of the spin Hall conductivity (SHC) components $\sigma_{xy}^z$ and $\sigma_{xz}^y$ at three different chemical potentials: $\mu = E_F$ (Fermi energy), $\mu = E_{SF}$ (Fermi energy shifted to four-fold crossing at $\Gamma$ point), and $\mu = E_{RF}$ (Fermi energy shifted to six-fold crossing at R-point). The results are summarized in Table~\ref{tab:table I}.

\begin{table}[h!]
\centering

\begin{tabular}{l@{\hskip 12pt}c@{\hskip 12pt}c@{\hskip 12pt}c}
\toprule
& \multicolumn{3}{c}{\textbf{SiTc Material}} \\
\cmidrule(lr){2-4}
\textbf{Chemical Potential} & $E_F$ & $E_{RF}$ & $E_{SF}$ \\
\midrule
$\sigma_{xy}^z$ & 52.25 & 105.44 & 258.37 \\
$\sigma_{xz}^y$ & -81.1 & -246.85 & -291.34 \\
\bottomrule
\end{tabular}
\caption{Spin Hall conductivity (SHC) components $\sigma_{xy}^z$ and $\sigma_{xz}^y$ (in $\hbar/e$ units) at different chemical potentials for \textit{SiTc}.}\label{tab:table I}
\end{table}

\begin{figure}[t]
\centering
    \includegraphics[width=1\linewidth, height=4.1cm]{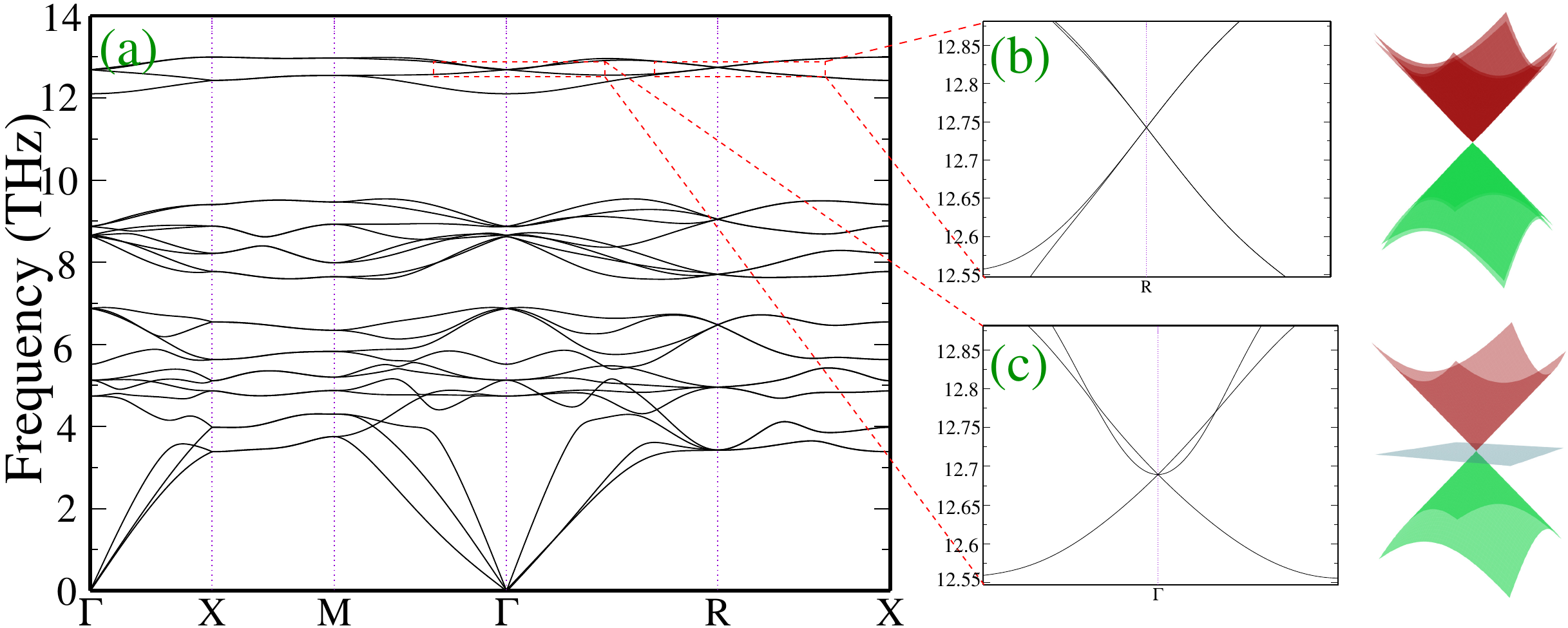}
    \caption{\label{fig:figure5} (a) Phonon band structure of the SiTc crystal, confirming the dynamical stability of the material. (b) Enlarged view of the three-fold bosonic band crossing at the $\Gamma$ high-symmetry point. (c) Four-fold bosonic band crossing at the R high-symmetry point, along with its 3D representation.}

\end{figure}

Phononic Topology :-    The SiTc contain 4 atoms of Si and 4 atoms of Tc shows 24 phonon frequency, mode where 3 lower frequency is acoustical and rest are optical modes. The effect of chiral crystal can also be seen in the phononic properties. The phonon band structure is calculated and shown in Figure \ref{fig:figure5}, has no imaginary phonon frequency indicating the dynamic stability of SiTc crystal structure. As the phononic system  do not follow the pauli exclusion principle, allow to show the topological properties to entire frequency range. Although by the symmetry analysis in section II of the chiral crystal discussed for without SOC of electronic properties can be completly justified for the phonon dispersion as the phonon does not effected by the spin orbital coupling, we have taken the upermost phonon bands in the range of 12 to 13 THz frequency range due to its well seprated bands. As we are intrested in the $\Gamma$ and R high symmetry point we have zoomed that part in the selected frequency range and observe that the $\Gamma$ possess 3-fold spin-1 excitation, whereas R-point possess 4-fold charge-2 Dirac point that are protected by $C_{3z}$ rotation symmetry, as shown in Figure \ref{fig:figure5}(b) and \ref{fig:figure5}(c) respectively.  

\begin{figure}[!t]
\includegraphics[width=1 \linewidth]{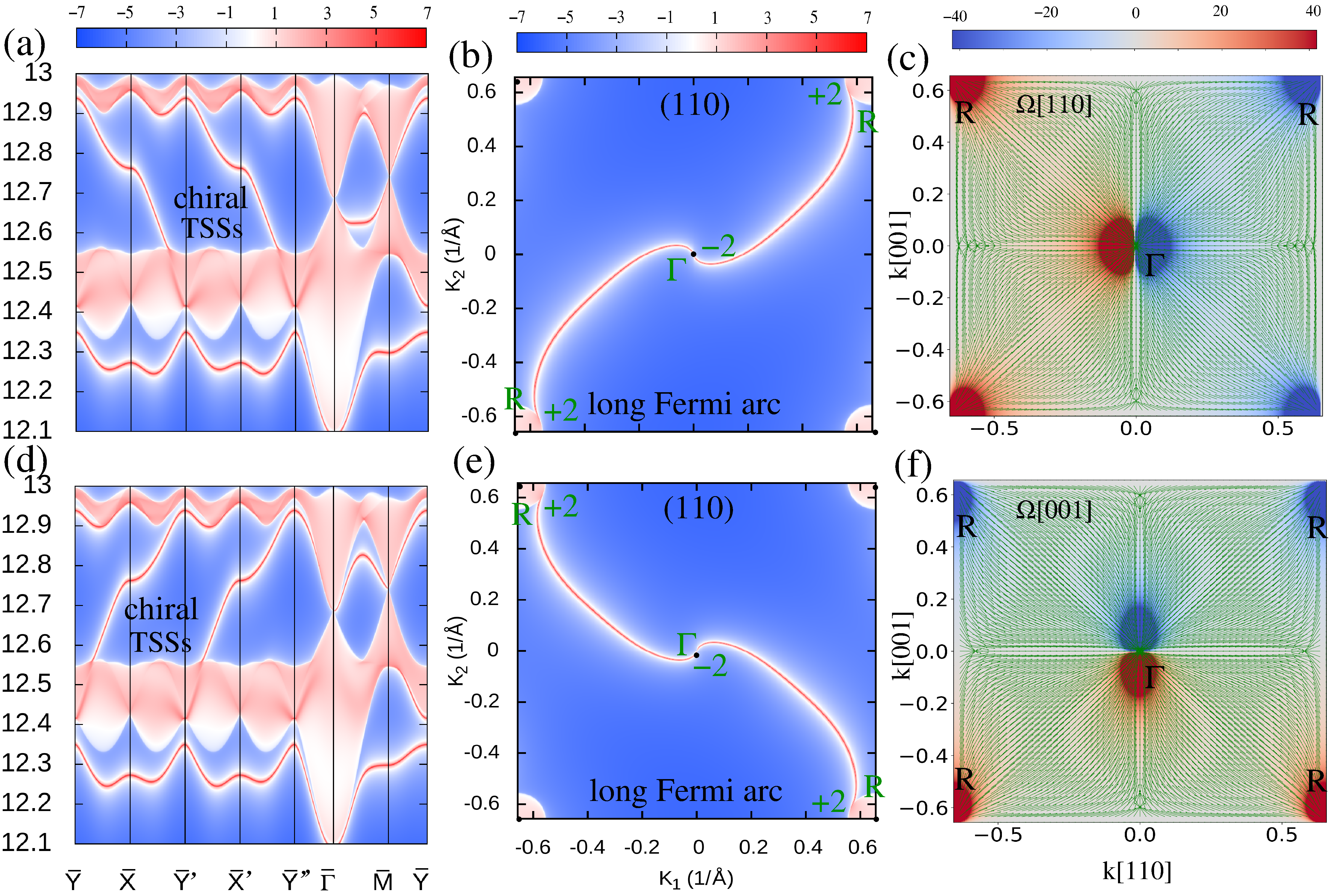}
 \caption{\label{fig:figure6} (a) Surface phonon band structure of the upper (001) surface, at local path, and (b) the corresponding Fermi surface, showing the longest Fermi arc connecting multifold phononic Weyl points of opposite chirality $\pm2$. (c) Distribution of Berry curvature $\Omega$ projected along the [110] direction, overlaid with the intensity of its in-plane (xy) and out-of-plane (z) components. Blue and red junctions indicate the presence of Weyl points. (d) Surface phonon band structure of the lower (001) surface, showing the continuation of chiral states from the upper surface, and (e) its corresponding Fermi surface, highlighting the longest chiral Fermi arc. (f) Distribution of Berry curvature $\Omega$ projected along the [001] direction, overlaid with the intensity of the xy and z components.}

 \end{figure}

To confirm the chirality and topological properties validation of SiTc, surface band structure with local symmetry path has been taken to show the chiral topological surface states that are completely mirror copy in there upper and lower surfaces as shown in Figure \ref{fig:figure6}(a) and \ref{fig:figure6}(d). The longest possible chiral open Fermi arc present that are connected to $\Gamma$(with -2 chirlality) and R (with +2 chirlality) high symmetry point in the first Brillouin zone, shown in Figure \ref{fig:figure6}(b) and \ref{fig:figure6}(e) . The complete open Fermi arc can be seen in one energy level is due to minimal energy window present in the system and the surfaces bands are well seperated from the bulk part unlike to the electronic band structure. Also the distribution of Berry curvature $\Omega$ overlaped with  intensity of the xy and z components of $\Omega$ has been calculated, shows the local vector fields and the junction of red (positive) and blue (negative) indicating the presence of  multifold Weyl point. The Berry curvature $\Omega$ on the $k_x-k_y$ plane flows almost directly from R to $\Gamma$ with minimal out-of-plane deviations, shown in Figure \ref{fig:figure6}(c) and \ref{fig:figure6}(f).

\section{Conclusion} 

The electronic and phononic properties of the chiral crystal SiTc have been systematically investigated using first-principles calculations combined with symmetry analysis. Our results confirm that SiTc hosts multifold topological features in both electronic and bosonic excitation spectra.
In the electronic structure, when spin–orbit coupling (SOC) is not considered, three-fold and four-fold degenerate fermionic states are identified at the $\Gamma$ and R high-symmetry points of the Brillouin zone, respectively. Upon incorporating SOC, these degeneracies transform into four-fold and six-fold fermions at the $\Gamma$ and R points, respectively. In addition to these symmetry-protected nodes, several Weyl points emerge along the $\Gamma$–R high-symmetry path and around the $\Gamma$ point. The total topological charge, obtained by summing the chiralities of all Weyl nodes within the Brillouin zone, vanishes, in agreement with the Nielsen–Ninomiya no-go theorem. Surface states and the corresponding Fermi arcs, along with the distribution of Berry curvature, have been calculated to support and visualize the topological nature of these band crossings. Furthermore, the intrinsic spin Hall conductivity (SHC) of SiTc has been computed, revealing that it originates from the large spin Berry curvature in the vicinity of the multifold fermionic nodes.
In the phonon dispersion, topological bosonic modes are observed, with three-fold and four-fold degeneracies located at the $\Gamma$ and R high-symmetry points, respectively. These bosonic nodes carry opposite chiralities ($\pm 2$) and are connected via longest possible surface phonon Fermi arcs, as evidenced by the surface phonon band structure and Berry curvature calculations.
The simultaneous presence of multifold topological features in both electronic and phononic excitations positions SiTc as a promising candidate for exploring new quantum phenomena. We expect that these theoretical predictions will inspire future experimental investigations to realize and probe these exotic quasiparticles in chiral crystals.

\section{Acknowledgments}
{This work was supported by the Anusandhan National Research Foundation (ANRF), Government of India, under Grant No. CRG/2022/006419, and by the Council of Scientific and Industrial Research–Human Resource Development Group (CSIR-HRDG) through the ASPIRE Grant No. 03WS(006)/2023-24/EMR-II/ASPIRE. The authors gratefully acknowledge the computational resources and infrastructure provided by Visvesvaraya National Institute of Technology (VNIT) Nagpur, which were essential to the completion of this study. PS acknowledges the high-performance computing support provided by the National Param Supercomputing Facility (NPSF) at C-DAC Pune. SKG acknowledges the financial support received through an institute fellowship from the Ministry of Human Resource Development (MHRD), Government of India. }

\bibliographystyle{ieeetr}
\bibliography{bib}


\end{document}